\documentclass[aps,prb,a4paper,twocolumn,floatfix,showkeys,amsmath,amssymb,nofootinbib,nobibnotes,altaffilletter]{revtex4}

\usepackage[T1]{fontenc}
\usepackage[latin1,utf8]{inputenc}
\usepackage[english]{babel}

\usepackage{setspace}

\usepackage{amsmath}
\usepackage{bm}

\usepackage{graphicx}
\usepackage{pslatex}
\usepackage{xspace}
\usepackage{subfigure}

\usepackage{booktabs}

\usepackage{ragged2e}
\usepackage[labelfont=bf, format=plain, indention=0.5cm, textfont=small, skip=4pt, tableposition=top, justification=RaggedRight, figureposition=bottom]{caption}

\usepackage{float} 
\floatstyle{boxed}

\begin{document}

\newfloat{copyrightfloat}{thp}{lop}
\begin{copyrightfloat}
\RaggedRight
The peer reviewed version of the following article has been published online on October 7, 2011, at physica status solidi (RRL), doi: 10.1002/pssr.201105430.
\end{copyrightfloat}

\preprint{}

\title{Photoinduced C$_{\textup{70}}$ radical anions in polymer:fullerene blends}

\title{Binding energy of singlet excitons and charge transfer complexes in MDMO-PPV:PCBM solar cells}

\author{Julia Kern$^1$}\email{julia.kern@physik.uni-wuerzburg.de}
\author{Sebastian Schwab$^1$}
\author{Carsten Deibel$^1$}
\author{Vladimir Dyakonov$^{1,2}$}

\affiliation{$^1$Experimental Physics VI, Julius-Maximilians University of W{\"u}rzburg, D-97074 W{\"u}rzburg}
\affiliation{$^2$Bavarian Center for Applied Energy Research (ZAE Bayern), 97074 W{\"u}rzburg, Germany}



\begin{abstract}
The influence of an external electric field on the photoluminescence intensity of singlet excitons and charge transfer complexes is investigated for a poly[2-methoxy-5-(3',7'-dimethyloctyloxy)-1,4-phenylenevinylene] (MDMO-PPV) diode and a bulk heterojunction of the PPV in combination with [6,6]-phenyl-C$_{61}$ butyric acid methylester (PCBM), respectively. The experimental data is related to the dissociation probability derived from the Onsager--Braun model. In this way, a lower limit for the singlet exciton binding energy of MDMO-PPV is determined as $(327 \pm 30)$ meV, whereas a significantly lower value of $(203 \pm 18)$ meV is extracted for the charge transfer complex in a MDMO-PPV:PCBM blend.
\end{abstract}



\maketitle

In recent years, joint efforts of a vast and expanding research community have lead to a steady improvement of organic solar cell (OSC) power conversion efficiencies. However, even OSC record efficiencies of currently 8.3 \% \cite{bib1} are still trailing behind the ones of their inorganic counterparts --- a fact which can be attributed to the relatively low dielectric constants ($\epsilon_r \approx 3 - 4$) of the employed organic semiconductors. The resulting weak screening causes the primary photoexcitations, i.e. the singlet excitons S$_1$ created in the polymer, to be strongly bound, thus making thermally induced dissociation to free charge carriers highly improbable. In the bulk heterojunction (BHJ) solar cell concept, dissociation is driven by ultrafast photoinduced charge transfer between a donor and an acceptor material (e.g. a conjugated polymer--fullerene system). The interpenetrated donor--acceptor network leads to the creation of charge transfer complexes (CTC), i.e. bound electron--hole pairs at the heterointerface forming after singlet exciton dissociation. While the energetics of those complexes are already known to influence the overall OSC efficiency by setting the maximum limit for the open circuit voltage \cite{bib5}, the extent of the CTC's impact on the equally important photocurrent is still controversially debated \cite{bib6,bib7}. Nevertheless, there exists a common notion that CTC can be precursors to at least some of the generated free charge carriers if the coulombic CTC binding energy can be overcome by an additional driving force. In order to contribute to the understanding of the underlying physical processes, we report on the determination of the binding energy $E_{b,S_1}$ of the singlet exciton in poly[2-methoxy-5-(3',7'-dimethyloctyloxy)-1,4-phenylenevinylene] (MDMO-PPV) and $E_{b,CTC}$ of the CTC of the PPV derivative blended with [6,6]-phenyl-C$_{61}$ butyric acid methylester (PCBM). 

Due to the finite binding energy, a certain amount of the created singlet excitons or CTC is not dissociated but recombines. The radiative part of the recombination can be observed as photoluminescence (PL). The application of an increasing external electric field constitutes an additional driving force which leads to an enhanced singlet exciton or CTC dissociation and thus to a reduction of the PL intensity. This field dependent PL quenching (PL(F)) has been observed for both singlet excitons and CTC for a variety of material systems \cite{bib9,bib20} and is also regarded as a suitable technique for determining the binding energy \cite{bib8,bib9} by relating it to the established Onsager--Braun model \cite{bib10,bib11}. This model comprises Braun's adaption of the Onsager theory to the case of the electron--hole pair being the lowest excited state of the observed system \cite{bib17} and thus applies to both the singlet exciton and the CTC. The respective dissociation probability is given by \cite{bib12}

\begin{equation}
\label{eq1}
P(F)=\frac{k_d(F)}{k_d(F)+k_f}=\frac{\kappa^*_d(F)}{\kappa^*_d(F)+(\mu\tau)^{-1}}
\end{equation}

\begin{figure}[h]%
\subfigure[Singlet exciton.]{%
\includegraphics*[width=.45\linewidth]{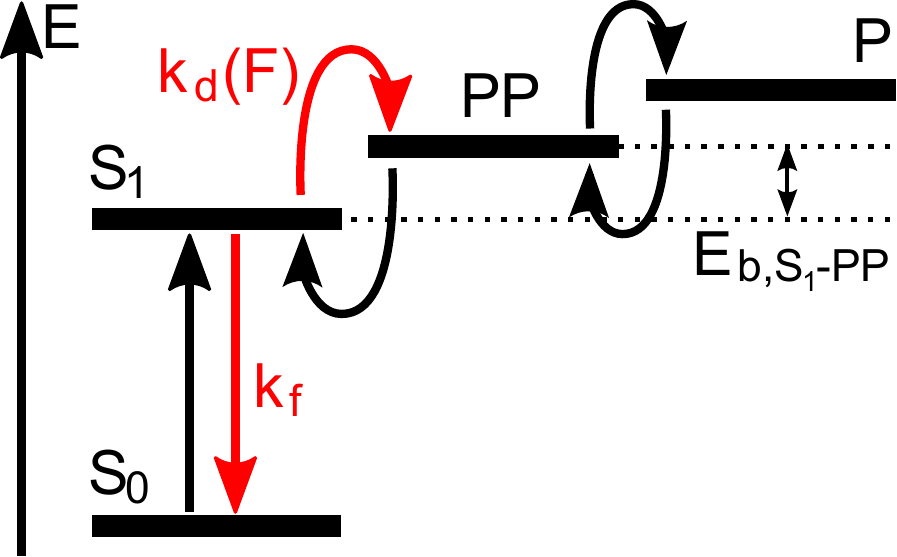}
\label{subfig1}
}\hfill
\subfigure[CTC.]{%
\includegraphics*[width=.45\linewidth]{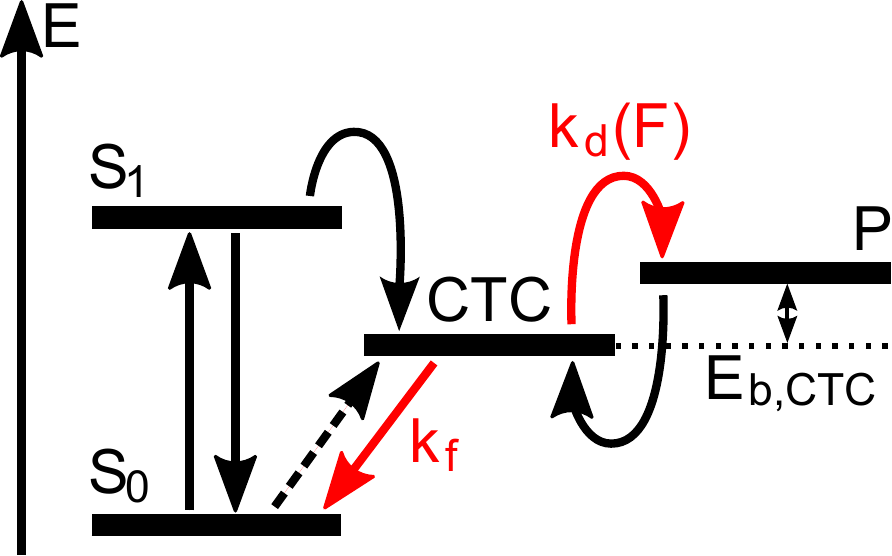}
\label{subfig2}}%
\caption{%
Rate models including the field dependent dissociation rate $k_d(F)$ and the singlet exciton or CTC recombination rate $k_f$ for a) MDMO-PPV and b) MDMO-PPV:PCBM.}
\label{rates}
\end{figure}

Here, $F$ is the electric field, $k_d(F)$ and $k_f$ are the singlet exciton or CTC dissociation and recombination rates, $\mu$ is the sum of electron and hole mobilities and $\tau=1/k_f$ is the lifetime. $k_d(F)=\mu\cdot\kappa^*_d(F)$ can be expressed as \cite{bib12}

\begin{equation}
\label{eq2}
k_d(F) = \frac {3 \gamma}{4 \pi r^3_s} \exp \left ( -\frac {E_b}{k_BT} \right ) \frac {J_1(2 \sqrt{-2b} ) } {\sqrt{-2b}}
\end{equation}

with the Langevin recombination factor $\gamma=q \mu / \epsilon_0 \epsilon_r$, the electron--hole distance $r_s$, the coulombic binding energy $E_b = e^2 / (4 \pi \epsilon_0 \epsilon_r r_s)$ and the thermal energy $k_BT$. $J_1$ is the Bessel function of order one and $b = e^3 F / \left (8 \pi \epsilon_0 \epsilon_r (k_BT)^2 \right )$ where $e$ is the elementary charge and $\epsilon_0 \epsilon_r$ is the effective dielectric constant in the bulk. The relevant rates for the dissociation of singlet excitons and CTC are shown in a simple energy diagram in Fig.~\ref{rates}.

As can be seen in Fig.~\ref{subfig1}, the PL(F) method can only give a lower limit for $E_{b,S_1}$ since the dissociation of the singlet exciton to polarons (P) involves a two-step process via the formation of a polaron pair (PP) \cite{bib9}. The dissociation probability given by the Onsager--Braun theory (Eq.~\ref{eq1}) can be related to the actual PL(F) measurements by calculating the PL quenching yield $\left [ PL(0)-PL(F) \right ] / PL(0)$, where $PL(F)$ and $PL(0)$ are the PL signals with and without an applied electric field, respectively.


For the experiments, a standard solar cell sample structure was utilised. A 35 nm layer of poly(3,4-ethylenedioxythiophene):poly(styrenesulfonate) was spin coated on top of an indium tin oxide covered glass substrate, followed by 60 -- 130 nm thick layers of the active materials (i.e. pure MDMO-PPV (Sigma-Aldrich) and MDMO-PPV blended with PCBM (Solenne) in a 1:1 weight ratio). Subsequently, the cathode consisting of Ca (3 nm) and Al (120 nm) was thermally evaporated.
Excitation was accomplished by a 532 nm laser. PL spectra were detected with a monochromator system whereas PL(F) for the determination of $E_{b}$ was recorded integrally by a stand-alone Si diode in combination with various long pass filters. Here, the cut-on wavelength was chosen as 550 nm for MDMO-PPV and 800 nm for MDMO-PPV:PCBM. An external voltage was applied in reverse bias direction in order to minimise effects of charge carrier injection. All experiments were conducted at room temperature.


\begin{figure}[h]%
\includegraphics*[width=1\linewidth]{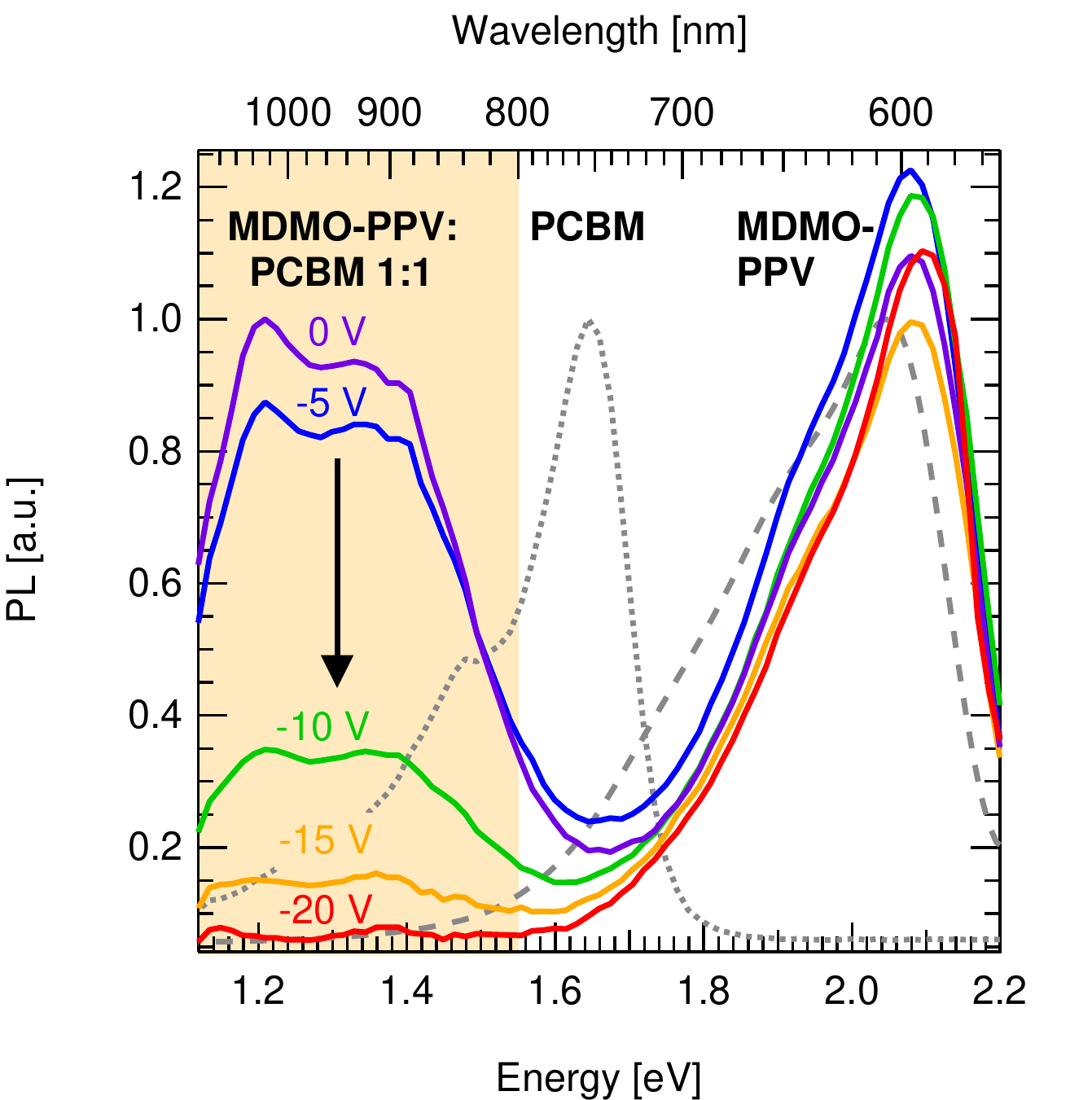}
\caption{%
PL spectra of MDMO-PPV (dashed line), PCBM (dotted line) and MDMO-PPV:PCBM in a 1:1 weight ratio (solid line labeled 0 V) and field dependent PL spectra of the blend system (solid lines). The highlighted rectangle marks the integrated spectral region for the PL(F) experiments on the blend system.}
\label{spectra}
\end{figure}

The observed PL spectra of MDMO-PPV, PCBM and the corresponding blend system are shown in Fig.~\ref{spectra}. The lower energetic peak in the blend spectrum is not apparent in any of the pure material spectra and is attributed to a CTC. It should be noted that for the chosen weight ratio of 1:1 -- unlike the in terms of photovoltaics better performing and therefore commonly used 1:4 ratio -- the blend spectrum does not show any residual PL originating from domains of pure PCBM in the bulk \cite{bib22}. Such an emission would cause an offset in the integral PL(F) signal and hinder the determination of $E_{b}$. 

\begin{figure}[h]%
\includegraphics*[width=1\linewidth]{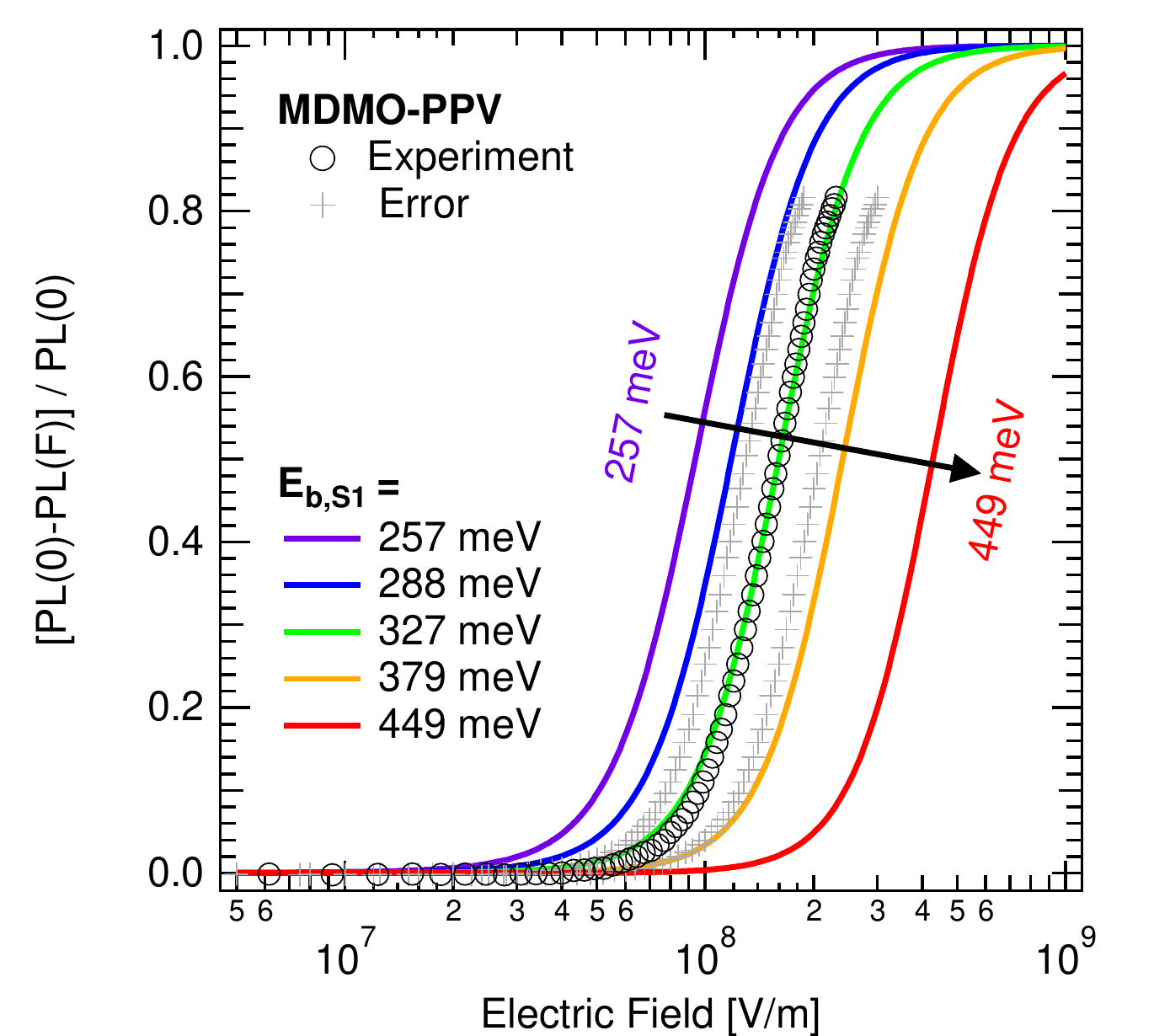}
\caption{%
PL quenching yield for the singlet exciton emission of MDMO-PPV (circles) in comparison to the dissociation probability predicted by the Onsager--Braun theory (solid lines).}
\label{EbS1}
\end{figure}
\begin{figure}[h]%
\includegraphics*[width=1\linewidth]{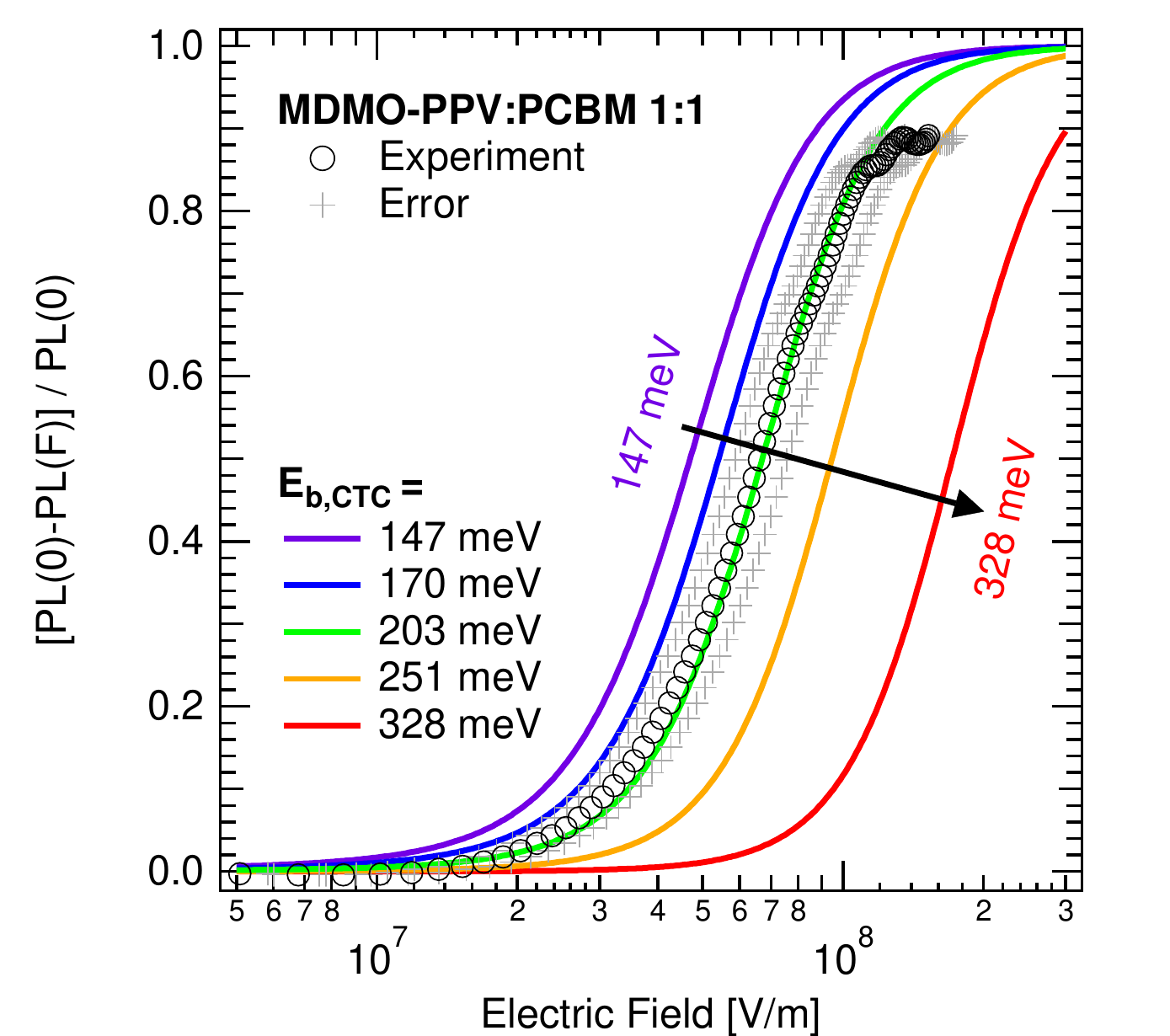}
\caption{%
PL quenching yield for the CTC emission of MDMO-PPV:PCBM (circles) in comparison to the dissociation probability predicted by the Onsager--Braun theory (solid lines).}
\label{EbCTE}
\end{figure}

The application of the Onsager--Braun model using the parameters $T = 297$ K, $\mu\tau = 1 \cdot 10^{-17}$ V/m and $\epsilon_r = 3.0$ for MDMO-PPV \cite{bib14} or $\epsilon_r = 3.4$ for MDMO-PPV:PCBM (with $\epsilon_r=3.9$ for PCBM \cite{bib15}) reveals a good agreement with the experimentally obtained data (see Fig.~\ref{EbS1} and \ref{EbCTE}). In contrast to other works \cite{bib13}, a high field dependent PL reduction of more than 80 \% is reached which allows for the accurate determination of $E_{b}$. The highest inaccuracy stems from measuring the sample layer thicknesses and thus from the calculation of the applied electric fields, leading to the curves showing the experimental error bars in Fig.~\ref{EbS1} and \ref{EbCTE}. For MDMO-PPV, a lower limit for the binding energy of the singlet exciton is determined as $E_{b,S_1} \ge (327 \pm 30)$ meV (Fig.~\ref{EbS1}), corresponding to an electron--hole distance of $r_s \approx  1.47$ nm. This $E_{b,S_1}$ is consistent with the 0.4 eV reported in literature for other PPV derivatives \cite{bib8,bib19}. 

In comparison, the binding energy of the CTC in MDMO-PPV:PCBM is found to be at a significantly lower value of $E_{b,CTC} = (203 \pm 18)$ meV (Fig.~\ref{EbCTE}) with an electron--hole distance of $r_s \approx 2.08$ nm. 

For a similar material system, Hallermann et al. \cite{bib13} reported a much smaller binding energy of $E_{b,CTC} = 130$ meV despite finding an only slightly different electron--hole distance $r_s = 1.88$ nm. However, the low binding energy implies an unrealistically high dielectric constant of $\epsilon_r = 5.9$ instead of the literature value of 3.4 \cite{bib14,bib15}. In any case, $E_{b,CTC}$ is found to be much higher than the thermal energy of about 25 meV which makes thermal dissociation unfeasible. CTC dissociation under solar cell working conditions (i.e. in the presence of electric fields below 10$^7$ V/m where
only a weak field dependence is observed \cite{bib23}) has recently been suggested to occur via excited (`hot') CTC \cite{bib6}, a mechanism whose relevance is supported by transient optical spectroscopy measurements on poly(3-hexylthiophene) (P3HT):PCBM devices \cite{bib21}. Yet, the CTC probed by PL in the presented work are more likely to be thermally relaxed so that the reported $E_{b,CTC}$ might only be an upper limit for the binding energy. These considerations emphasise the need for further investigations concerning this topic in order to elucidate the detailed mechanism for charge generation. 


In conclusion, we have related field dependent PL measurements to the established Onsager--Braun model. In this way, we were able to accurately determine a lower limit for the binding energy of the singlet excitons in MDMO-PPV ($E_{b,S_1} \ge (327 \pm 30)$ meV) and an upper limit for the binding energy of the CTC in MDMO-PPV:PCBM ($E_{b,CTC} \le (203 \pm 18)$ meV).

 \hspace{5pt}


\begin{acknowledgements}
The current work is supported by the Bundesministerium für Bildung und Forschung in the framework of the GREKOS project (contract no. 03SF0356B) and the Deutsche Forschungsgemeinschaft in the framework of the SPP1355 project Phorce (DE 830/8-1). C.D. gratefully acknowledges the support of the Bavarian Academy of Sciences and Humanities. V.D.'s work at the ZAE Bayern is financed by the Bavarian Ministry of Economic Affairs, Infrastructure, Transport and Technology.
\end{acknowledgements}

\end{document}